# Computing the wavefunction from trajectories: particle and wave pictures in quantum mechanics and their relation

**Peter Holland**

Green College
University of Oxford
Woodstock Road
Oxford OX2 6HG
England

peter.holland@green.ox.ac.uk

20 May 2004

## Abstract

We discuss the particle method in quantum mechanics which provides an exact scheme to calculate the time-dependent wavefunction from a single-valued continuum of trajectories where two spacetime points are linked by at most a single orbit. A natural language for the theory is offered by the hydrodynamic analogy, in which wave mechanics corresponds to the Eulerian picture and the particle theory to the Lagrangian picture. The Lagrangian model for the quantum fluid may be developed from a variational principle. The Euler-Lagrange equations imply a fourth-order nonlinear partial differential equation to calculate the trajectories of the fluid particles as functions of their initial coordinates using as input the initial wavefunction. The admissible solutions are those consistent with quasi-potential flow. The effect of the superposition principle is represented via a nonclassical force on each particle. The wavefunction is computed via the standard map between the Lagrangian coordinates and the Eulerian fields, which provides the analogue in this model of Huygens principle in wave mechanics. The method is illustrated by calculating the time-dependence of a free Gaussian wavefunction. The Eulerian and Lagrangian pictures are complementary descriptions of a quantum process in that they have associated Hamiltonian formulations that are connected by a canonical transformation. The de Broglie-Bohm interpretation, which employs the same set of trajectories, should not be conflated with the Lagrangian version of the hydrodynamic interpretation. The theory implies that the mathematical results of the de Broglie-Bohm model may be regarded as statements about quantum mechanics itself rather than about its interpretation.

# 1. Introduction

The notion that the concept of a continuous material orbit is incompatible with a full wave theory of microphysical systems was central to the genesis of wave mechanics [1]. Early attempts to justify this assertion using Heisenberg's relations were subsequently shown to be flawed, and indeed no credible proof forbidding the treatment of quantum processes in terms of precisely defined spacetime trajectories has ever been offered. The idea nevertheless entered the folklore of the subject and even now is invoked to highlight alleged paradoxical implications of quantum mechanics (e.g., Schrödinger's cat). So great was the philosophical bias that not only was the material orbit ruled out as an aid to comprehension, but the possibilities of using the trajectory as a computational tool, or even as the basis of an alternative representation of the quantum theory - the twin subjects of this paper – were foregone.

In the path-integral method the path concept is integral to the computation of the wavefunction: the propagator linking two spacetime points is calculated by linearly superposing the elementary amplitudes associated with all the paths connecting the points [2]. The wavefunction at a point is then found from Huygens principle by superposing the contributions coming from all the other points (weighted by the initial wavefunction). The technique gives the impression that the trajectory concept works in this context as a computational device only because it is allied with a simultaneous application of the superposition principle to the path. Thus, the propagation of a particle somehow involves the simultaneous traversal of multiple equally likely paths. The method therefore rather reinforces the view that a description of the propagation between two points using a *single* spacetime track is incompatible with the quantum description of the process.

Of course, de Broglie and Bohm demonstrated long ago the falseness of this conclusion – no rule of quantum theory is broken by employing a single trajectory to specify the state of a system more finely than does the wavefunction [3, 4]. But, being just an interpretative element rather than an intrinsic ingredient of quantum theory, it has been possible to dismiss the de Broglie-Bohm trajectory as a "superfluous 'ideological superstructure'" [5]. Indeed, the very method of computing the trajectory, which relies on first knowing the wavefunction whose determination is independent of the path, appears to confirm the de Broglie-Bohm trajectory as a redundant addendum.

The question whether the concept of a single path linking two spacetime points, such as that employed by de Broglie and Bohm, can be made the basis of an alternative computational scheme in quantum mechanics has received scant attention in the foundations of physics literature. Indeed, it appears to be unknown that the seeds of an alternative view of the de Broglie-Bohm trajectories were sown more than 30 years ago when it was shown that they could be instrumental in solving numerically the time-dependent Schrödinger equation [6-9]. Rather than being merely interpretative, the trajectories acquired a use-value. Further work was done on this "particle (or trajectory) method" [10, 11] and recently it has flourished, principally in the chemical physics literature, with several significant developments in the technique being reported (for a selection of references see [12-33]).

The approach arises from the well-known analogy between quantum mechanics and hydrodynamics. The similarity in form between the single-body Schrödinger equation in the position representation (wave mechanics), suitably expressed as two real equations, and the equations of fluid mechanics was first pointed out by Madelung [34]. In this analogy, the probability density is proportional to the fluid density and the phase of the wavefunction is a velocity potential. A novel feature of the quantum fluid is the appearance of quantum stresses, usually represented through the quantum potential. To achieve full mathematical equivalence of the models, the hydrodynamic variables have to satisfy conditions inherited from the wavefunction. These in turn provide physical insight into the original conditions. For example, the single-valuedness requirement on the wavefunction corresponds to the appearance of quantized vortices in the fluid [35-37].

The analogy prompts the following observation. For a classical fluid one can adopt the Eulerian picture, in which one observes the flow from fixed space points and to which the Schrödinger fluid-like equations correspond in the analogy just mentioned, or the Lagrangian picture, in which one follows individual fluid particles. The idea explored in the numerical work cited above is to adapt elements of the Lagrangian method of the classical theory to quantum hydrodynamics as a means of solving the wave equation. For example, the fluid may be approximated by a finite collection of representative particles and their equations of motion solved simultaneously with those of the phase and amplitude functions to generate expressions for the wave function along each trajectory. These techniques generally have advantages over methods employing just the Eulerian equations for the computational effort is naturally concentrated in regions where the density of trajectories is highest.

The aim of this paper is to elaborate on the general theory of the Lagrangian method in quantum mechanics in order to bring out its full significance. The Lagrangian model for the quantum fluid may be developed systematically from (variational) first principles, in much the way one proceeds in classical hydrodynamics (Sec. 2). The Euler-Lagrange equations provide an exact formula (a version of Newton's second law) to calculate the trajectories of the fluid particles as functions of their initial coordinates using as input the initial wavefunction. The latter data restricts the admissible solutions to those consistent with quasi-potential flow (Sec. 3). Using these solutions we obtain a quantum analogue of the classical construction of the time-dependent Eulerian fluid functions purely from the fluid particle trajectories (Sec. 4). The key step in the computation is to express the initial coordinates as functions of space and time. One thereby achieves the same end as the path-integral method – computation of the wavefunction given its initial value - using a quite different and conceptually simpler trajectory method in which two spacetime points are connected by at most a single continuous orbit. The two steps in Feynman's approach (propagator, then Huygens) are thus condensed into one, the wavefunction as a whole being generated from its initial form by a single-valued continuum of trajectories. The superposition principle enters at the level of the path not in the sense that many different routes through space are available to a particle starting from one point but indirectly in that on the unique path that the particle takes between two points it is subject to nonclassical effects due to an additional quantum term in Newton's force law. However, unlike the path-integral method this is not so far a method of "quantization" since the functional form of the fluid internal potential energy, the origin of the quantum effects, is unexplained.

The fact that we can derive the time-dependent wavefunction from the trajectories (an example is given in Sec. 5) indicates that we are dealing here not just with an interpretation (hydrodynamic), but rather with an alternative mathematical representation, or picture, of quantum mechanics. The hydrodynamic analogy is a particularly suggestive and natural conceptual resource for presenting the theory but the mathematical development is prompted by rather than reliant upon it.

The full hydrodynamic model of quantum mechanics therefore provides an interpretation of two pictures – the wave-mechanical (Eulerian) and the particle (Lagrangian), and the latter is just as valid a representation of quantum processes as the former. The word "picture" seems appropriate, for we shall see that the two versions of quantum mechanics have associated Hamiltonian formulations that are connected by a canonical transformation, much as the familiar pictures of quantum mechanics (e.g., Schrödinger and Heisenberg) are connected by unitary transformations (Sec. 6). This mapping therefore gives a new and mathematically precise meaning to the notion of "wave-particle duality". Indeed, the Lagrangian particle coordinates and their canonical momenta have the character of $c$-number analogues of Heisenberg-picture operators.

Without going into the intricacies of what it means to "interpret" quantum mechanics [38], it is emphasized that, although it is closely related mathematically, and is cited as inspiration for the numerical studies mentioned above, the de Broglie-Bohm interpretation should not be conflated with the Lagrangian version of the hydrodynamic interpretation. In particular, it is necessary to distinguish the de Broglie-Bohm corpuscle from a fluid particle (Sec. 7).

The work we report is a simple generalization and combination of several strands of classical Lagrangian fluid theory, which came to fruition in the 1980s. Background material is given by Lamb [39] and the history and elements of the theory are recorded in the beautiful book of Truesdell [40] (see also [41]). Classical variational treatments are described in [42-56] and relevant work on the hydrodynamic approach to quantum mechanics in [3, 35-37, 57-67]. A variational treatment similar to that presented here has been given by Rylov [68] who has considered a generalization of the classical theory to include dependence of the internal energy on higher derivatives of the density of the kind needed in the quantum case, and has written down a special case of one version of the equation of motion for the fluid coordinates. Rylov approaches the Lagrangian theory in a somewhat different spirit, regarding it as descriptive of an ensemble of stochastic processes, and does not discuss the benefits of the approach in providing an alternative method of computation, or other aspects such as the canonical mapping. Other examples of the mapping between Eulerian field equations and Lagrangian coordinates are discussed in [69].

## 2. Lagrangian picture

The Lagrangian picture in hydrodynamics is a field theory but its discourse is drawn from particle mechanics. A fluid is a continuum of elements (particles) distributed throughout Euclidean space, each of which retains its identity throughout the evolution of the whole. The history of the system is encoded in the state variables $q(a,t)$, the positions of all the distinct fluid elements at time $t$, each particle being distinguished by a continuously-variable vector label $a$. For convenience we take $a$ to be the particle position at $t = 0$. The motion is *continuous* in that the mapping from $a$-space to $q$-space is single-valued and differentiable with respect to $a$ and $t$ to whatever order is necessary, and the inverse mapping $a(q,t)$ exists and has the same properties. These assumptions are in accord with the properties of the velocity field implied by quantum mechanics (i.e., the ratio of current to density), in particulate its single-valuedness. The entire set of motions for all $a$ is termed a *flow*. The vectors $q$ and $a$ are referred to the same set of Cartesian space axes but they may also be regarded as related to one another by a time-dependent coordinate transformation. To each particle there is associated an elementary volume whose mass is conserved by the flow. The whole is structured by an internal potential derived from the density that represents a certain kind of particle interaction, and each particle responds to the potential via a force whose action is described by a form of Newton's second law. For all these reasons, we shall regard the model as providing a "particle" picture.

Let $\rho_0(a)$ be the initial quantal probability density. In the hydrodynamic model $\rho_0(a)$ is identified with the initial number density (which is normalized: $\int \rho_0(a)d^3a = 1$). Then, introducing a mass parameter $m$, the mass of an elementary volume $d^3a$ attached to the point $a$ is given by $m\rho_0(a)d^3a$. The significance of the parameter $m$, conventionally described as the "mass of the quantum system", is that it is the total mass of the fluid, since $\int m\rho_0(a)d^3a = m$. In this picture, the conservation of the mass of a fluid element in the course of its motion is expressed through the relation

$$m\rho(q(a,t))d^3q(a,t) = m\rho_0(a)d^3a \qquad (2.1)$$

or

$$\rho(a,t) = J^{-1}(a,t)\rho_0(a) \tag{2.2}$$

where $J$ is the Jacobian of the transformation between the two sets of coordinates:

$$J = \frac{1}{3!}\varepsilon_{ijk}\varepsilon_{lmn}\frac{\partial q_i}{\partial a_l}\frac{\partial q_j}{\partial a_m}\frac{\partial q_k}{\partial a_n}, \quad 0 < J < \infty. \tag{2.3}$$

Here $\varepsilon_{ijk}$ is the completely antisymmetric tensor with $\varepsilon_{123} = 1$ and the indices $i,j,k,\ldots = 1,2,3$. Summation over repeated indices is always assumed.

Let $V$ be the potential of an external (classical) conservative body force and $U$ the internal potential energy of the fluid due to interparticle interactions. We assume that the Lagrangian has the same form as in the classical theory of ideal fluids except for the functional dependence of $U$: this depends on $\rho(q)$ and its first derivatives, and hence from (2.2) on the second-order derivatives of $q$ with respect to $a$, and is independent of other variables such as entropy. The Lagrangian is then

$$\begin{aligned}
L[q,\partial q/\partial t,t] &= \int l\big(q,\partial q/\partial t,\partial q/\partial a,\partial^2 q/\partial a^2,t\big)d^3a \\
&= \int\Big[\tfrac{1}{2}m\rho_0(a)\big(\partial q(a,t)/\partial t\big)^2 - \rho_0(a)U(\rho) - \rho_0(a)V\big(q(a)\big)\Big]d^3a.
\end{aligned} \tag{2.4}$$

Here $\rho_0(a)$ and $V$ are prescribed functions and we substitute for $\rho$ from (2.2). We assume that $\rho_0$ and its derivatives vanish at infinity, which ensures that the surface terms in the variational principle vanish. If in addition to the equations for the coordinates $q$ we wished to derive (2.2) from a variational principle we could add $\lambda(J\rho - \rho_0)$ to the integrand in (2.4), where $\lambda$ is an undetermined multiplier, and vary independently with respect to $\lambda$, $q$ and $\rho$. This construction is unnecessary for our purposes here (we shall however absorb the (Eulerian version of) the conservation equation into a canonically-related action principle in Sec. 6).

It is the action of the conservative force derived from $U$ on the trajectories that represents the quantum effects in this theory. As we shall see, these effects are characterised by the following choice for $U$ (motivated by the well known Eulerian expression for the internal energy):

$$U = \frac{\hbar^2}{8m}\frac{1}{\rho^2}\frac{\partial\rho}{\partial q_i}\frac{\partial\rho}{\partial q_i} = \frac{\hbar^2}{8m}\frac{1}{\rho_0{}^2}J_{ij}J_{ik}\frac{\partial}{\partial a_j}\left(\frac{\rho_0}{J}\right)\frac{\partial}{\partial a_k}\left(\frac{\rho_0}{J}\right) \tag{2.5}$$

where we have substituted from (2.2) and used

$$\frac{\partial}{\partial q_i} = J^{-1}J_{ij}\frac{\partial}{\partial a_j} \tag{2.6}$$

where

$$J_{il} = \frac{\partial J}{\partial(\partial q_i/\partial a_l)} = \frac{1}{2}\varepsilon_{ijk}\varepsilon_{lmn}\frac{\partial q_j}{\partial a_m}\frac{\partial q_k}{\partial a_n} \tag{2.7}$$

is the cofactor of $\partial q_i/\partial a_l$. The latter satisfies

$$\frac{\partial q_k}{\partial a_j} J_{ki} = J \delta_{ij}.$$ (2.8)

Clearly, $U$ has a local dependence on $\rho$ and its derivatives and the coordinates $q$ enter only through the deformation gradients $\partial q_i / \partial a_j$ and their derivatives with respect to $a$. This dependence implies an interaction between each fluid particle and its neighbours. The fact of interaction in the quantum case does not signal a conceptual departure from classical fluid dynamics where likewise the internal forces depend on the deformation gradients. The difference here is that the order of the derivative-coupling of the particles is higher than is customary in a classical equation of state, and implies a considerably more complex and subtle mutual particle dependence. Other features of the quantum flow, such as non-crossing of the paths, are also anticipated in the behaviour of classical ideal fluids but note that not all features are – there is a notable difference in the vortex structures, for example (see Sec. 3).

The Euler-Lagrange equations for the coordinates,

$$\frac{\partial}{\partial t}\frac{\partial L}{\partial (\partial q_i(a,t)/\partial t)} - \frac{\delta L}{\delta q_i(a)} = 0$$ (2.9)

where

$$\frac{\delta L}{\delta q_i} = \frac{\partial l}{\partial q_i} - \frac{\partial}{\partial a_j}\frac{\partial l}{\partial (\partial q_i/\partial a_j)} + \frac{\partial^2}{\partial a_j \partial a_k}\frac{\partial l}{\partial (\partial^2 q_i/\partial a_j \partial a_k)},$$ (2.10)

give

$$m\rho_0(a)\frac{\partial^2 q_i(a)}{\partial t^2} = -\rho_0(a)\frac{\partial V}{\partial q_i} - \frac{\partial W_{ij}}{\partial a_j}$$ (2.11)

where

$$W_{ij} = -\rho_0(a)\frac{\partial U}{\partial (\partial q_i/\partial a_j)} + \frac{\partial}{\partial a_k}\left(\rho_0(a)\frac{\partial U}{\partial (\partial^2 q_i/\partial a_j \partial a_k)}\right)$$ (2.12)

They obviously have the form of Newton's second law. We shall not give the explicit form of the quantity $W_{ij}$ but will utilise instead a more useful related tensor, $\sigma_{ij}$, defined by $W_{ik} = J_{jk}\sigma_{ij}$. The latter is the analogue in this theory of the classical pressure tensor, $p\delta_{ij}$. Using (2.8) we can invert to obtain

$$\sigma_{ij} = J^{-1}W_{ik}\frac{\partial q_j}{\partial a_k}.$$ (2.13)

Evaluating (2.12) and using the relation (2.8) and its derivatives with respect to $\partial q_m / \partial a_n$ the pressure tensor may be written

$$\sigma_{ij} = \frac{\hbar^2}{4mJ^3} J_{ik} \left[ \rho_0^{-1} J_{jl} \frac{\partial \rho_0}{\partial a_k} \frac{\partial \rho_0}{\partial a_l} + \left( J^{-1} J_{jl} J_{mn} - J_{jm\,ln} \right) \frac{\partial \rho_0}{\partial a_l} \frac{\partial^2 q_m}{\partial a_k \partial a_n} - J_{jl} \frac{\partial^2 \rho_0}{\partial a_k \partial a_l} \right.$$
$$\left. + \rho_0 \left( J^{-1} J_{mn} J_{jrls} + J^{-1} J_{jl} J_{mrns} - 2J^{-2} J_{jl} J_{mn} J_{rs} \right) \frac{\partial^2 q_r}{\partial a_k \partial a_s} \frac{\partial^2 q_m}{\partial a_l \partial a_n} + \rho_0 J^{-1} J_{jl} J_{mn} \frac{\partial^3 q_m}{\partial a_k \partial a_l \partial a_n} \right] \tag{2.14}$$

where

$$J_{jm\,ln} = \frac{\partial J_{jl}}{\partial (\partial q_m / \partial a_n)} = \varepsilon_{jmk} \varepsilon_{ln\,r} \frac{\partial q_k}{\partial a_r}. \tag{2.15}$$

It is easy to check that this tensor is symmetric. The equation of motion of the $a$th fluid particle moving in the "field" of the other particles and the external force is now

$$m\rho_0(a) \frac{\partial^2 q_i(a)}{\partial t^2} = -\rho_0(a) \frac{\partial V}{\partial q_i} - J_{kj} \frac{\partial \sigma_{ik}}{\partial a_j} \tag{2.16}$$

where we have used the identity

$$\frac{\partial J_{ij}}{\partial a_j} = 0. \tag{2.17}$$

This (second order in $t$, fourth order in $a$) local nonlinear partial differential equation (or its equivalent forms (2.18), (3.6) and (3.11) below) is the principal analytical result of the quantum Lagrangian method. For we shall see that from its solutions $q_i(a,t)$, subject to specification of $\partial q_{i0}/\partial t$ whose determination is discussed in the next section, we may derive solutions to Schrödinger's equation. The remainder of the paper is mainly concerned with the various facets of the relationship between the two dynamical equations.

Other forms of (2.16) will be useful. Multiplying by $\partial q_i / \partial a_k$ we obtain the "Lagrangian" form:

$$m\rho_0(a) \frac{\partial^2 q_i(a)}{\partial t^2} \frac{\partial q_i}{\partial a_k} = -\rho_0(a) \frac{\partial V}{\partial a_k} - \frac{\partial q_i}{\partial a_k} J_{kj} \frac{\partial \sigma_{ik}}{\partial a_j}. \tag{2.18}$$

In addition, it is straightforward to develop the Hamiltonian formalism for the quantum fluid. The canonical field momenta are

$$p_i(a) = \frac{\partial L}{\partial (\partial q_i(a)/\partial t)} = m\rho_0(a) \frac{\partial q_i(a)}{\partial t} \tag{2.19}$$

so that, making the usual Legendre transformation, the Hamiltonian is

$$H[q,p,t] = \int p_i(a) \frac{\partial q_i(a)}{\partial t} d^3 a - L[q, \partial q/\partial t, t]$$
$$= \int \left[ \frac{p(a)^2}{2m\rho_0(a)} + \rho_0(a) U\left( J^{-1}\rho_0 \right) + \rho_0(a) V\left( q(a) \right) \right] d^3 a. \tag{2.20}$$

Hamilton's equations can be written in terms of the usual Poisson brackets, where the basic relations are

$$\{q_i(a), q_j(a')\} = \{p_i(a), p_j(a')\} = 0, \quad \{q_i(a), p_j(a')\} = \delta_{ij}\delta(a - a').$$ (2.21)

We obtain

$$\frac{\partial q_i(a)}{\partial t} = \frac{\delta H}{\delta p_i(a)}, \quad \frac{\partial p_i(a)}{\partial t} = -\frac{\delta H}{\delta q_i(a)}$$ (2.22)

which when combined reproduce (2.16).

## 3. Quasi-potential flow

To obtain a flow that is representative of quantum mechanics we need to restrict the initial conditions of (2.16) to those that correspond to what we shall term "quasi-potential" flow. This means that the initial velocity field is of the form (we introduce the mass factor for later convenience)

$$\frac{\partial q_{i0}(a)}{\partial t} = \frac{1}{m}\frac{\partial S_0(a)}{\partial a_i}$$ (3.1)

but the flow is not irrotational everywhere because the potential $S_0(a)$ (the initial quantal phase) obeys the quantization condition

$$\oint_C \frac{\partial q_{i0}(a)}{\partial t} da_i = \oint_C \frac{1}{m}\frac{\partial S_0(a)}{\partial a_i} da_i = \frac{nh}{m}, \quad n \in Z,$$ (3.2)

where $C$ is a closed curve composed of material particles. If it exists, vorticity occurs in nodal regions (where the density vanishes) and it is assumed that $C$ passes through a region of "good" fluid, where $\rho_0 \neq 0$. To show that these assumptions imply motion characteristic of quantum mechanics we first demonstrate that they are preserved by the dynamical equation. To this end, we use the method based on Weber's transformation applied to the law of motion in the Lagrangian form (2.18) [39, 40] together with a quantum analogue of Kelvin's circulation theorem. Other methods are available, such as Cauchy's vorticity formula [40], but we shall not discuss them here.

We first get (2.18) into a more convenient form. Expressed in terms of the dependent variables $q(a)$, and using (2.2), the stress tensor (2.14) takes a simpler form:

$$\sigma_{ij} = \frac{\hbar^2}{4m}\left(\frac{1}{\rho}\frac{\partial \rho}{\partial q_i}\frac{\partial \rho}{\partial q_j} - \frac{\partial^2 \rho}{\partial q_i \partial q_j}\right)$$ (3.3)

Using the easily proved but important result

$$\frac{1}{\rho}\frac{\partial \sigma_{ij}}{\partial q_j} = \frac{\partial V_Q}{\partial q_i}$$ (3.4)

where

$$V_Q = \frac{\hbar^2}{4m\rho}\left(\frac{1}{2\rho}\frac{\partial\rho}{\partial q_i}\frac{\partial\rho}{\partial q_i} - \frac{\partial^2\rho}{\partial q_i\partial q_i}\right)$$ (3.5)

is the de Broglie-Bohm quantum potential [3, 4][1], we see that (2.18) may also be simplified:

$$m\frac{\partial^2 q_i}{\partial t^2}\frac{\partial q_i}{\partial a_k} = -\frac{\partial}{\partial a_k}(V + V_Q).$$ (3.6)

We now integrate this equation between the time limits $(0,t)$:

$$m\frac{\partial q_i}{\partial t}\frac{\partial q_i}{\partial a_k} = m\frac{\partial q_{k0}}{\partial t} + \frac{\partial\chi(a,t)}{\partial a_k}$$ (3.7)

where

$$\chi(a,t) = \int_0^t\left(\frac{1}{2}m\left(\frac{\partial q}{\partial t}\right)^2 - V - V_Q\right)dt.$$ (3.8)

Then, substituting (3.1),

$$\frac{\partial q_i}{\partial t}\frac{\partial q_i}{\partial a_k} = \frac{1}{m}\frac{\partial S}{\partial a_k}, \quad S(a,t) = S_0(a) + \chi(a,t),$$ (3.9)

with initial conditions $q = a$, $\chi_0 = 0$. The left-hand side of (3.9) gives the velocity at time $t$ with respect to the $a$-coordinates and this is obviously a gradient. To obtain the $q$-components we multiply by $J^{-1}J_{ik}$ and use (2.6) and (2.8) to get

$$\frac{\partial q_i}{\partial t} = \frac{1}{m}\frac{\partial S}{\partial q_i}$$ (3.10)

where $S = S(a(q,t),t)$. Thus, for all time the velocity of each particle is the gradient of a potential with respect to the current position.

Eq. (3.10) is a form of the law of motion. Corresponding to this, it is notable that we may write (3.6) in terms of the current variables as well:

$$m\frac{\partial^2 q_i}{\partial t^2} = -\frac{\partial}{\partial q_i}(V + V_Q).$$ (3.11)

This puts the fluid-dynamical law of motion (2.16) in the form of Newton's law for a particle of mass $m$ moving in the potential $V + V_Q$, a point we discuss in Sec. 7. In the case where we substitute (2.2) in (3.5), a special case of (3.11) (corresponding to $\rho_0 = 1$) has been written down by Rylov [68].

---

[1] Madelung was one of the first to recognize the importance of this quantity [34]. The earliest use of the expression "quantum potential" is apparently due to de Broglie [70].

To complete the demonstration, we note that the motion is quasi-potential since the value (3.2) of the circulation is preserved following the flow:

$$\frac{\partial}{\partial t} \oint_{C(t)} \frac{\partial q_i}{\partial t} \, dq_i = 0 \tag{3.12}$$

where $C(t)$ is the evolute of the material particles that compose $C$. This theorem has been stated previously in the quantum context [35] and is proved in the usual way [44, 71]. We conclude that each particle retains forever the quasi-potential property if it possesses it at any moment.

To obtain the equation obeyed by the potential $S$, we use the chain rule applied to a function $F(a,t)$ with $a = a(q, t)$:

$$\left.\frac{\partial F}{\partial t}\right|_a = \left.\frac{\partial F}{\partial t}\right|_q + \frac{\partial q_i}{\partial t} \frac{\partial F}{\partial q_i}. \tag{3.13}$$

Then, since from (3.9) $\chi = S - S_0$,

$$\left.\frac{\partial \chi}{\partial t}\right|_a = \left.\frac{\partial S}{\partial t}\right|_q + \frac{\partial q_i}{\partial t} \frac{\partial S}{\partial q_i} - \left( \left.\frac{\partial S_0}{\partial t}\right|_q + \frac{\partial q_i}{\partial t} \frac{\partial S_0}{\partial q_i} \right) \tag{3.14}$$

Using (3.13) the two terms in the brackets sum to $\partial S_0(a)/\partial t = 0$. Using (3.10), (3.14) therefore becomes

$$\left.\frac{\partial \chi}{\partial t}\right|_a = \left.\frac{\partial S}{\partial t}\right|_q + m\left(\frac{\partial q}{\partial t}\right)^2. \tag{3.15}$$

Thus, since from (3.8)

$$\left.\frac{\partial \chi}{\partial t}\right|_a = \frac{1}{2} m\left(\frac{\partial q}{\partial t}\right)^2 - V - V_Q, \tag{3.16}$$

we obtain finally, using (3.10),

$$\frac{\partial S}{\partial t} + \frac{1}{2m}\left(\frac{\partial S}{\partial q}\right)^2 + V + V_Q = 0. \tag{3.17}$$

This is the "quantum Hamilton-Jacobi equation"[2], one half of Schrödinger's equation, which will be derived again in Sec. 4. We have therefore shown that the set of equations comprising the conservation law (2.2) and the Euler-Lagrange equations (2.16), together with the initial condition (3.1), are equivalent to the five equations (2.2), (3.9) and (3.17). These equations serve to determine the functions $q_i, \rho, S$, which are now the unknowns. The completion of the demonstration that the collective particle motion generates Schrödinger evolution is given in Sec. 4. Note that although the particle velocity is orthogonal to a moving surface $S = \text{constant}$, the surface does not keep step with the particles that initially compose it and hence it is not a material surface.

---

[2] The validity of this epithet is discussed in [72].

There is a noteworthy difference between quantum and classical vortex lines. One of the classical Helmholtz laws states that vortex lines are material lines, i.e., they are composed of the same material particles for all time, and so move with the fluid [73]. In contrast, in the quantum case vorticity occurs only where $\rho = 0$, i.e., where there are no material particles, so the vortex lines (strictly, line vortices) cannot be material lines. Indeed, $\partial q/\partial t = m^{-1}\partial S/\partial q$ is singular at nodes so the material velocity does not give a description of the vortex motion. Rather, the spacetime behaviour is given directly by the (two real) condition(s) $\psi(x,t) = 0$ (in Eulerian terms; see Sec. 4). The vortex motion is nevertheless closely related to the surrounding local flow, as is obvious from the circulation theorem (3.12). Thus, the loop $C(t)$ can be chosen arbitrarily small as long as it does not pass through a node. Then a vortex line initially inside $C(t)$ cannot cross it because this would imply a change in the circulation, contradicting the theorem. Therefore, the vortex is trapped inside, and moves "with", the surrounding "good" fluid. The quantum analogue of the classical theorem is therefore that vortex lines move with the fluid, but are not material lines. The motion of quantum vortices is discussed in [3, 36, 37, 66, 67].

## 4. Eulerian picture. Schrödinger's equation

The fundamental link between the particle (Lagrangian) and wave-mechanical (Eulerian) pictures is defined by the following expression for the Eulerian density:

$$\rho(x,t) = \int \delta\big(x - q(a,t)\big)\rho_0(a)\,d^3a. \tag{4.1}$$

This relation may be regarded as a component of a transformation from the three coordinate functions $q_i(a)$ to a new system of coordinate functions one of which is $\rho(x)$ – indeed, as we shall see in Sec. 6, (4.1) is a building block of the canonical transformation linking the two pictures. The corresponding formula for the Eulerian velocity $v$ is contained in the expression for the current:

$$\rho(x,t)v_i(x,t) = \int \frac{\partial q_i(a,t)}{\partial t}\delta\big(x - q(a,t)\big)\rho_0(a)\,d^3a. \tag{4.2}$$

Relations (4.1) and (4.2) play an analogous role in our approach to the Huygens formula

$$\psi(x,t) = \int G(x,t;a,0)\psi_0(a)\,d^3a \tag{4.3}$$

in Feynman's method, for from them we can construct the wavefunction, up to a global phase. In view of the similar role it plays to $G$ in (4.3), we may refer to the delta function $\delta\big(x - q(a,t)\big)$ as the "propagator" of the motion in time. Unlike the many-to-many mapping embodied in the formula (4.3), the quantum evolution is described here by a local point-to-point development. Using the result

$$\delta\big(x - q(a,t)\big) = J^{-1}\big|_{a(x,t)}\,\delta\big(a - a_0(x,t)\big), \quad x - q(a_0,t) = 0, \tag{4.4}$$

and evaluating the integrals, (4.1) and (4.2) are equivalent to the following local expressions (we give also the inverse relations)

$$\rho(x,t) = J^{-1}\big|_{a(x,t)}\rho_0\big(a(x,t)\big), \quad \rho\big(x(a,t),t\big) = J^{-1}(a,t)\rho_0(a) \tag{4.5}$$

$$v_i(x,t) = \frac{\partial q_i(a,t)}{\partial t}\bigg|_{a(x,t)}, \quad v_i(x,t)\big|_{x(a,t)} = \frac{\partial q_i(a,t)}{\partial t}. \tag{4.6}$$

Relations (4.5) restate the conservation equation (2.2), and (4.6) give the relations between the velocities in the two pictures. We may term $J^{-1}$ the "local propagator". Note that, unlike the Feynman propagator, the fluid propagator, which is determined by solutions to (2.16), depends on $\psi_0(a)$.

From the first relation (4.6) we may deduce the following relations between the accelerations in the two pictures:

$$\frac{\partial v_i}{\partial t} + v_j \frac{\partial v_i}{\partial x_j} = \frac{\partial^2 q_i(a,t)}{\partial t^2}\bigg|_{a(x,t)}, \quad \left(\frac{\partial v_i}{\partial t} + v_j \frac{\partial v_i}{\partial x_j}\right)_{x(a,t)} = \frac{\partial^2 q_i(a,t)}{\partial t^2}. \tag{4.7}$$

These formulas enable us to translate the Lagrangian flow equations into Eulerian language. Differentiating (4.1) with respect to $t$ and using (4.2) we easily deduce the continuity equation

$$\frac{\partial \rho}{\partial t} + \frac{\partial}{\partial x_i}(\rho v_i) = 0. \tag{4.8}$$

Next, differentiating (4.2) and using (3.11) and (4.8) we get the quantum analogue of Euler's equation:

$$\frac{\partial v_i}{\partial t} + v_j \frac{\partial v_i}{\partial x_j} = -\frac{1}{m}\frac{\partial}{\partial x_i}(V + V_Q). \tag{4.9}$$

Finally, the quasi-potential condition (3.10) becomes

$$v_i = \frac{1}{m}\frac{\partial S(x,t)}{\partial x_i}. \tag{4.10}$$

The first members in (4.5) and (4.6) give the general solutions of the continuity equation (4.8) and Euler's equation (4.9), respectively, in terms of the paths and initial density.

To establish the connection between the Eulerian equations and Schrödinger's equation, we note that, using (4.10), (4.9) can be written

$$\frac{\partial}{\partial x_i}\left(\frac{\partial S}{\partial t} + \frac{1}{2m}\frac{\partial S}{\partial x_i}\frac{\partial S}{\partial x_i} + V + V_Q\right) = 0. \tag{4.11}$$

The quantity in brackets is thus a function of time. Since the addition of a function of time to $S$ does not affect the velocity field, we may absorb the function in $S$, i.e., set it to zero. Then

$$\frac{\partial S}{\partial t} + \frac{1}{2m}\frac{\partial S}{\partial x_i}\frac{\partial S}{\partial x_i} + V + V_Q = 0. \tag{4.12}$$

Combining (4.8), (4.10) and (4.12), the function $\psi(x,t) = \sqrt{\rho}\exp(iS/\hbar)$ obeys Schrödinger's equation:

$$i\hbar\frac{\partial\psi}{\partial t} = -\frac{\hbar^2}{2m}\frac{\partial^2\psi}{\partial x_i\partial x_i} + V\psi. \tag{4.13}$$

We have deduced this from the Lagrangian particle equation (2.16) subject to the quasi-potential requirement. The quantization condition (3.12) becomes here

$$\oint_{C_0}\frac{\partial S(x,t)}{\partial x_i}dx_i = nh, \quad n \in Z, \tag{4.14}$$

where $C_0$ is a closed curve fixed in space that does not pass through nodes. This is a consistent subsidiary condition on solutions since it is easy to see that the value of (4.14) is preserved in time as long as nodes do not cross $C_0$ (an Eulerian version of the circulation theorem).

Given the initial wavefunction $\psi_0(a) = \sqrt{\rho_0}\exp(iS_0/\hbar)$ we can compute the wavefunction for all $x,t$, up to a global phase, as follows. First, solve (2.16) subject to the initial conditions $q_0(a) = a$, $\partial q_{i0}(a)/\partial t = m^{-1}\partial S_0(a)/\partial a$ to get the set of trajectories for all $a,t$. Next, substitute $q(a,t)$ in (4.5) to find $\rho$ and $\partial q/\partial t$ in (4.6) to get $\partial S/\partial x$. This gives $S$ up to an additive function of time, $f(t)$. To fix this function, apart from an additive constant, use (4.12). We obtain finally the following formula for the wavefunction:

$$\psi(x,t) = \sqrt{\left(J^{-1}\rho_0\right)\big|_{a(x,t)}}\exp\left[\frac{i}{\hbar}\left(\int m\,\partial q_i(a,t)/\partial t\big|_{a(x,t)}dx_i + f(t)\right)\right] \tag{4.15}$$

Note the key role played by the initial coordinates in this formula.

The Eulerian equations (4.8) and (4.9) form a closed system of four first-order coupled partial differential equations to determine the four independent fields $\rho(x)$, $v_i(x)$ and do not refer to the material paths. We shall call them the "basic" Eulerian equations. The erasure of the particle variables is part of the reason why the Eulerian language is particularly suited to represent the wave-mechanical formalism which likewise, of course, makes no reference to the trajectory concept (another reason is that classical hydrodynamics already exhibits the key feature of mutual coupling between the density and velocity transport equations). It will be noted, however, that the Lagrangian theory from which we derived the Eulerian system comprises seven independent fields $\rho$, $q(a)$, $p(a)$. (In the case of quasi-potential flow there are two (five) independent Eulerian (Lagrangian) fields.) Depending on the information one wants about the physical system this may be regarded as indicating a redundancy in the Lagrangian description, or an incompleteness in the Eulerian one. In the sense that the Eulerian view refers only to the variables of interest, namely, the wavefunction, then the Lagrangian method may seem profligate. The situation is similar to that of the Feynman paths where likewise the bones of the calculus are washed out in the final answer. On the other hand, from the perspective of establishing the mathematical equivalence of the pictures we must adopt the second option. The necessity of supplementing the Eulerian picture with variables that do not appear in the basic equations is well known in the classical variational theory [45-56].

A simple solution is to append to the Eulerian description the three particle-label functions $a_i(x,t)$. Their equations of evolution state that the labels are conserved by the flow implied by the Eulerian velocity field:

$$\frac{\partial a_i}{\partial t} + v_j\frac{\partial a_i}{\partial x_j} = 0. \tag{4.16}$$

Additional structure of this type arises naturally in the canonical theory described in Sec. 6. Eq. (4.16) is equivalent to the second version of the equation of motion (4.6). To see this, multiply (4.16) by $\partial x_k / \partial a_i$ to get

$$v_j = -\frac{\partial x_k}{\partial a_i}\frac{\partial a_i}{\partial t}. \tag{4.17}$$

Differentiating the relation $x\left(a(x,t),t\right) = x$ with respect to $t$ we therefore obtain

$$v_i(x,t)\big|_{x(a,t)} = \frac{\partial x_i(a,t)}{\partial t}. \tag{4.18}$$

Starting from the Eulerian approach the trajectories become derived quantities (which we emphasize by using the label $x$ in place of $q$). We can pass to the Lagrangian picture using the conversion formulas (4.5) and (4.6) above.

## 5. Example

The special case of a one-dimensional flow may be extracted from the three-dimensional theory when $V(x) = V_1(x_1) + V_2(x_2) + V_3(x_3)$, for solutions to the Eulerian equations exist for which $\rho(x) = \rho_1(x_1)\rho_2(x_2)\rho_3(x_3)$ and $S(x) = S_1(x_1) + S_2(x_2) + S_3(x_3)$. In these circumstances the internal potential decomposes:

$$U = \frac{\hbar^2}{8m}\left(\left(\frac{1}{\rho_1}\frac{\partial \rho_1}{\partial q_1}\right)^2 + \left(\frac{1}{\rho_2}\frac{\partial \rho_2}{\partial q_2}\right)^2 + \left(\frac{1}{\rho_3}\frac{\partial \rho_3}{\partial q_3}\right)^2\right) \tag{5.1}$$

In the Lagrangian picture, (3.10) implies that this condition translates as independence of the orthogonal coordinates: $q_1 = q_1(a_1,t)$ etc. Eq. (2.16) for $q_1$, say, then becomes, dropping the index,

$$m\rho_0\frac{\partial^2 q}{\partial t^2} = -\rho_0\frac{\partial V}{\partial q} + \frac{\hbar^2}{4m}\frac{\partial}{\partial a}\left(2\rho_0 J^{-5}\left(\frac{\partial J}{\partial a}\right)^2 - J^{-4}\frac{\partial J}{\partial a}\frac{\partial \rho_0}{\partial a} - \rho_0 J^{-4}\frac{\partial^2 J}{\partial a^2} + J^{-3}\frac{\partial^2 \rho_0}{\partial a^2} - J^{-3}\frac{1}{\rho_0}\left(\frac{\partial \rho_0}{\partial a}\right)^2\right) \tag{5.2}$$

with $J = \partial q / \partial a$.

To illustrate the method, we use formula (5.2) to compute the free ($V = 0$) wavefunction at time $t$ that is initially a Gaussian at rest:

$$\rho_0(a) = \left(2\pi\sigma_0^2\right)^{-1/2} e^{-a^2/2\sigma_0^2}, \quad S_0(a) = 0. \tag{5.3}$$

Let us try a solution of the form $q(a,t) = A(a)T(t)$. The initial conditions $q_0(a) = a$, $\partial q_0(a)/\partial t = 0$ imply

$$A = a, T_0 = 1, \partial T_0/\partial t = 0 \quad (a \neq 0)$$
$$A = 0, \text{ or } T_0 = \partial T_0/\partial t = 0, \text{ or both} \quad (a = 0). \Biggr\} \tag{5.4}$$

When $a \neq 0$, (5.2) becomes, noting that $J = T$ and hence $\partial J/\partial a = 0$,

$$m\rho_0 a \frac{\partial^2 T}{\partial t^2} = \frac{\hbar^2}{4mT^3} \frac{\partial}{\partial a}\left(\frac{\partial^2 \rho_0}{\partial a^2} - \frac{1}{\rho_0}\left(\frac{\partial \rho_0}{\partial a}\right)^2\right) \tag{5.5}$$

Substituting from (5.3), this reduces to

$$\frac{\partial^2 T}{\partial t^2} = \frac{\alpha}{T^3}, \quad \alpha = \left(\frac{\hbar}{2m\sigma_0{}^2}\right)^2. \tag{5.6}$$

The solution of (5.6) respecting the stated initial conditions is $T = \left(1 + \alpha t^2\right)^{1/2}$ and the paths are therefore

$$q(a, t) = a\left(1 + \alpha t^2\right)^{1/2}. \tag{5.7}$$

When $a = 0$, $A = 0$ implies $q = 0$ for all $t$. When $A \neq 0$, it is easy to see that the only solution obeying $T_0 = \partial T_0/\partial t = 0$ is $T(t) = 0$ so that again $q = 0$. Hence, (5.7) holds for all $a$.

The wavefunction at time $t$ may now be found by substituting (5.7) into the one-dimensional analogues of (4.5) and (4.6) (with (4.10)):

$$\rho(x, t) = \left(\partial q/\partial a\right)^{-1}\Big|_{a(x,t)} \rho_0\big(a(x, t)\big) \tag{5.8}$$

$$\partial S(x, t)/\partial x = m\,\partial q(a, t)/\partial t\big|_{a(x,t)}. \tag{5.9}$$

We find

$$\rho(x, t) = \left(2\pi\sigma^2\right)^{-1/2} e^{-x^2/2\sigma^2}, \quad S(x, t) = \frac{1}{2}m\alpha t x^2 \sigma_0{}^2 \sigma^{-2} - \frac{\hbar}{2}\tan^{-1}\left(\frac{\hbar t}{2m\sigma_0{}^2}\right) \tag{5.10}$$

where $\sigma = \sigma_0\left(1 + \alpha t^2\right)^{1/2}$ and the time-dependent addition in $S$ is obtained by putting (5.9) in (4.12). This is the correct time-dependent free Gaussian wavefunction.

## 6. Canonical transformation linking the particle and wave pictures

The absence of the material paths in the basic Eulerian equations suggests that $\rho(x)$ and $S(x)$ may be regarded as "collective coordinates" – functions that describe the bulk properties of the system without depending on the complex details of the particulate substructure. Indeed, (4.1) is a continuum analogue of the microscopic particle density of discrete particle mechanics,

$$\rho(x, t) = \sum_a \delta\big(x - q_a(t)\big), \tag{6.1}$$

which in certain contexts [74-76] is a collective coordinate. This idea has been examined previously in the classical fluid-dynamical context [77, 78].

The notion that $\rho(x)$ and $S(x)$ should be regarded as (canonical) coordinates for each $x$ is of course the basis of the standard variational treatment of the Schrödinger equation (where the alternative canonical variables $\psi, \psi*$ are commonly used; see below), and the coordinate interpretation finds its most natural expression in that context. Here we shall show how to pass from the Hamiltonian theory of the Lagrangian picture to a corresponding Hamiltonian theory of the Eulerian picture via a suitably chosen canonical transformation in which $q_i(a), p_i(a)$ are the "old" phase space coordinates and $\rho(x), S(x)$ are related to a "new" set of phase space coordinates denoted by $Q_i(x), P_i(x)$.

Apart from applying the quasi-potential restriction, the quantum treatment follows closely the corresponding classical analysis for vortical flows given by van Saarloos [52]. We suppose that the generating function of the canonical transformation is a time-independent functional of the old coordinates and the new momenta: $W[q(a), P(x)]$. The transformation formulas are therefore

$$Q_i(x) = \frac{\delta W}{\delta P_i(x)}, \quad p_i(a) = \frac{\delta W}{\delta q_i(a)}, \quad K[Q, P, t] = H[q, p, t].$$ (6.2)

The $W$ we seek is a generalization to continuous variables of the generating function of an arbitrary coordinate transformation:

$$W = \int \delta(x - q(a)) \rho_0(a) (P_1(x) + A(a)P_2(x) + B(a)P_3(x)) d^3a \, d^3x.$$ (6.3)

Here the functions $A$ and $B$ are arbitrary (e.g., we may choose $A = a_2, B = a_3$) subject to certain conditions such as that the Jacobian of the transformation is unity. Our transformation equations are then

$$Q_1(x) = \int \delta(x - q(a)) \rho_0(a) \, d^3a = (J^{-1}\rho_0)\big|_{a(x)} = \rho(x),$$
$$Q_2(x) = \rho(x)A(a(x)), \quad Q_3(x) = \rho(x)B(a(x)),$$ (6.4)

where we have substituted $\rho$ using (4.5), and

$$p_i(a) = \rho_0(a) \left( \frac{\partial P_1(x)}{\partial x_i} + A(a) \frac{\partial P_2(x)}{\partial x_i} + B(a) \frac{\partial P_3(x)}{\partial x_i} \right)_{x(a)}.$$ (6.5)

The transformation is therefore of the type sought in which $\rho$ becomes a coordinate. Combining these formulas and substituting in (2.20) the new Hamiltonian is

$$K[Q, P, t] = \int \left[ \frac{Q_1(x)}{2m} \left( \frac{\partial P_1(x)}{\partial x} + \frac{Q_2(x)}{Q_1(x)} \frac{\partial P_2(x)}{\partial x} + \frac{Q_3(x)}{Q_1(x)} \frac{\partial P_3(x)}{\partial x} \right)^2 + Q_1(x)U(Q_1, \nabla Q_1) + Q_1(x)V(x) \right] d^3x.$$ (6.6)

Using (2.19) and (4.6) the expression (6.5) for the momentum is just an extended Clebsch representation of the Eulerian velocity (any vector field can be expressed in this form [54]):

$$mv_i(x) = \frac{\partial P_1(x)}{\partial x_i} + A\big(a(x)\big)\frac{\partial P_2(x)}{\partial x_i} + B\big(a(x)\big)\frac{\partial P_3(x)}{\partial x_i}. \tag{6.7}$$

Using this shorthand along with $Q_1 = \rho$ the Hamiltonian can be written in terms of the non-canonical variables $\rho$, $v$, as in classical hydrodynamics:

$$K = \int \left(\tfrac{1}{2}m\rho v^2 + \rho U(\rho, \nabla\rho) + \rho V\right) d^3 x. \tag{6.8}$$

Hamilton's equations for the new variables are

$$\frac{\partial Q_1}{\partial t} = \frac{\delta K}{\delta P_1} = -\frac{\partial}{\partial x_j}\big(Q_1 v_j\big) \tag{6.9}$$

$$\frac{\partial Q_2}{\partial t} = \frac{\delta K}{\delta P_2} = -\frac{\partial}{\partial x_j}\big(Q_2 v_j\big) \tag{6.10}$$

$$\frac{\partial Q_3}{\partial t} = \frac{\delta K}{\delta P_3} = -\frac{\partial}{\partial x_j}\big(Q_3 v_j\big) \tag{6.11}$$

$$\frac{\partial P_1}{\partial t} = -\frac{\delta K}{\delta Q_1} = \frac{1}{2}mv^2 - v_i\frac{\partial P_1}{\partial x_i} - V_Q - V, \qquad V_Q = \frac{\delta}{\delta Q_1}\int Q_1 U(Q_1, \nabla Q_1)\, d^3 x \tag{6.12}$$

$$\frac{\partial P_2}{\partial t} = -\frac{\delta K}{\delta Q_2} = -v_i\frac{\partial P_2}{\partial x_i} \tag{6.13}$$

$$\frac{\partial P_3}{\partial t} = -\frac{\delta K}{\delta Q_3} = -v_i\frac{\partial P_3}{\partial x_i}. \tag{6.14}$$

It is convenient to replace the equations for $Q_2$ and $Q_3$ by equations obeyed by the functions $A$ and $B$ given in (6.4):

$$\frac{\partial A}{\partial t} + v_i\frac{\partial A}{\partial x_i} = 0, \quad \frac{\partial B}{\partial t} + v_i\frac{\partial B}{\partial x_i} = 0. \tag{6.15}$$

Hamilton's equations therefore become (6.9) and (6.12) together with the four equations (6.13)-(6.15) which state that each of the functions $A, B, P_2, P_3$ is a constant following the flow. Eq. (6.9) is just the Eulerian conservation equation (4.8) and it is straightforward to derive Euler's equation (4.9) from (6.7) using (6.12)-(6.15).

The coordinates $P_i$ are defined by relation (6.7) only up to a certain transformation, an analogue of the corresponding result for the usual Clebsch or Euler representation of a vector field [42, 44, 79]. For, suppose (6.7) is satisfied by another set of parameters $A', B', P_1', P_2', P_3'$. Then, subtracting the new relation from the old, we have

$$0 = \frac{\partial(P_1 - P_1')}{\partial x_i} + A\frac{\partial P_2}{\partial x_i} - A'\frac{\partial P_2'}{\partial x_i} + B\frac{\partial P_3}{\partial x_i} - B'\frac{\partial P_3'}{\partial x_i}. \tag{6.16}$$

This differential relation implies that

$$P_1' - P_1 = f, \quad A = \frac{\partial f}{\partial P_2}, \quad B = \frac{\partial f}{\partial P_3}, \quad A' = -\frac{\partial f}{\partial P_2'}, \quad B' = -\frac{\partial f}{\partial P_3'} \qquad (6.17)$$

for some function $f\left(P_2, P_2', P_3, P_3'\right)$. The variables $A, B, P_2, P_3$ and $A', B', P_2', P_3'$ are therefore connected by a canonical transformation generated by $f$. The functions $A, B$ being fixed parameters of the transformation (6.3), we have $A = A', B = B'$ and the generating function has the form $f\left(P_2 - P_2', P_3 - P_3'\right)$.

To complete the canonical mapping linking the pictures we must state the conditions obeyed by the new variables corresponding to the initial quasi-potential requirement in the old variables. At time $t$, the latter condition reads

$$\frac{p_i(a)}{\rho_0(a)} \frac{\partial q_i}{\partial a_k} = \frac{\partial S}{\partial a_k} \qquad (6.18)$$

(obtained by combining (2.19) with (3.9)). The relations (6.18) imply that the initial data for Hamilton's equations is given by the four independent functions $S_0, q_{i0}(= a_i)$ ( $\rho_0$ being regarded as a prescribed function). From (4.10) we have initially $m v_{i0} = \partial S_0 / \partial x_i$ which is connected with the initial variables $Q_{i0}, P_{i0}$ via (6.7). At time $t$ we then have

$$\frac{\partial}{\partial x_i}\left(S - P_1\right) = A \frac{\partial P_2}{\partial x_i} + B \frac{\partial P_3}{\partial x_i}. \qquad (6.19)$$

These relations fix the functions $A$ and $B$ in terms of the gradients of $S$ and $P_i$. Following (6.17), (6.19) implies that

$$S - P_1 = g, \quad A = \frac{\partial g}{\partial P_2}, \quad B = \frac{\partial g}{\partial P_3} \qquad (6.20)$$

for some function $g\left(P_2, P_3\right)$. This establishes the required relations between the $Q$s and $P$s. To confirm this, we derive the equation obeyed by $S$. Thus, substituting for $P_1$ in (6.12) and using (6.13) and (6.14) we obtain (4.12), as required. In addition, equations (6.13) and (6.14) imply (6.15) via the second and third relations in (6.20). The initial data is therefore given by $S_0, Q_{10}(= \rho_0), P_{20}, P_{30}$.

We conclude that Hamilton's equations (6.9)–(6.14) for $Q, P$, subject to the restrictions (6.20) imposed at $t = 0$, imply the two Eulerian equations that make up Schrödinger's equation together with two equations determining constants of the motion.

Some words of clarification on the canonical theory may be useful. In the usual variational approach to the Schrödinger equation the Hamiltonian is written

$$K'[\psi, \psi*] = \int \left( \frac{\hbar^2}{2m} \nabla \psi * . \nabla \psi + V|\psi|^2 \right) d^3 x \qquad (6.21)$$

where $\psi, \psi*$ are conjugate variables. Alternatively, we may make a canonical transformation to the conjugate variables $\rho(= Q'), S(= P')$ and get the Hamiltonian

$$K''[\rho, S] = \int \rho \left( \frac{1}{2m} (\nabla S)^2 + U(\rho, \nabla \rho) + V \right) d^3x. \tag{6.22}$$

Hamilton's equations for $Q'$, $P'$ lead to the required self-contained equations for $\rho, S$. The point of the demonstration above is that we can achieve the same set of self-governing equations using an extended Hamiltonian formalism where $\rho$ is one of three coordinate functions and $S$ enters not as its conjugate quantity but through an initial functional relationship that is maintained by Hamilton's equations. It is the enhanced canonical version of wave mechanics, rather than the conventional one, that maps to the particle Hamiltonian theory.

Finally, we give the Lagrangian in the Eulerian picture:

$$
\begin{aligned}
L[P, \partial P/\partial t, t] &= \int Q_i(x) \frac{\partial P_i(x)}{\partial t} d^3x - K[Q, P, t] \\
&= \int \rho \left( \frac{\partial P_1}{\partial t} + A \frac{\partial P_2}{\partial t} + B \frac{\partial P_3}{\partial t} - \tfrac{1}{2} mv^2 - U(\rho, \nabla \rho) - V \right) d^3x.
\end{aligned}
\tag{6.23}
$$

Naturally, the Euler-Lagrange equations for $P_i, \rho, A$ and $B$ reproduce Hamilton's equations (6.9)–(6.14), after some rearrangement.

## 7. Relation with the de Broglie-Bohm theory and related hydrodynamic approaches

It will be noted that the law of motion (4.18) (or (3.11)) for the fluid elements coincides with that postulated in the de Broglie-Bohm interpretation of spin 0 quantum mechanics [3, 4]. It is important, however, to distinguish the present approach from the general issue of interpretation, and to discriminate its hydrodynamic interpretation from the de Broglie-Bohm interpretation. This is particularly important as the fluid and de Broglie-Bohm approaches are sometimes discussed as if they are identical.

Regarding the general problem, we emphasize that, in the first instance, the purpose of the present theory is to provide an alternative method of computation that, in itself, is neutral regarding interpretation (in this regard it has a status analogous to that of the path-integral method). Of course, this mathematical formulation is particularly suited to a hydrodynamic discourse and hence may be regarded as having a bearing on the meaning of quantum theory. In this model, the fluid is composed of a continuum of fictitious "probability elements" each propagating deterministically along a fluid path (we have used "mass elements" in order to present the theory as fully as possible in the hydrodynamic language). The probability here refers to the distribution of measurement outcomes. The mass of a fluid particle is fixed by the initial density on the relevant path. All the fluid particles are present together and their collective motion is an alternative representation of quantum evolution. Unlike the original Madelung (Eulerian) formulation, the Lagrangian picture adds variables ($a$) to the quantum formalism but none is singled out as special.

In contrast, we recall that the purpose of the de Broglie-Bohm approach is to underpin the quantum statistical description with a theory of individual causally connected physical processes, a problem that is not addressed by the hydrodynamic approach in its primitive form. For this purpose, a corpuscle of mass $m$ is introduced and the wavefunction (strictly, gauge-invariant functions of it [72]) is attributed ontological status[3]. The statistical aspect of the wavefunction is now a secondary property and has a different meaning in referring to the likely current location of the particle rather than just the measured location. Although mathematically similar, the de Broglie-Bohm theory is logically independent of and involves elements additional to the hydrodynamic model. Within the context of the hydrodynamic model, the de Broglie-Bohm corpuscle should be regarded as a passive foreign body immersed in the fluid, or "tracer", a lone fluid path being singled out as its propagation track[4]. The mass point may lie on any path for which the probability is finite but the continuum of paths is realized only over an (infinite) ensemble of trials in each of which just one is traversed. In the original form of the theory, the law of motion of the corpuscle was appended to the Schrödinger equation in its Eulerian form but it can equally well be appended to the Lagrangian form (where it coincides with that of one of the fluid particles). *The de Broglie-Bohm theory involves a further level of interpretation of quantum mechanics beyond the hydrodynamic interpretation; this includes both the Eulerian and Lagrangian pictures* (or any other formulation) and it should not be identified with either. We attribute deviations in the motion of the corpuscle to local action of the quantum potential in the former case and to interactions between fluid particles in the latter; in both cases the corpuscle is a passive entity that does not act back upon the guiding agent. The corpuscle is therefore to be distinguished from a fluid particle both in its mass and in its dynamical behaviour. That the de Broglie-Bohm approach relates to a different *physical* problem to the hydrodynamic model becomes clear when we consider, for example, the phase space theory of the field-corpuscle system where it is necessary to introduce contributions for both the Schrödinger field (or set of fluid elements) and the corpuscle in the total Hamiltonian in order to achieve a consistent theory of their interaction [72].

---

[3] It is sometimes overlooked that these features do not fully characterise the de Broglie-Bohm model, or rather, it does not fulfil its potential as a physical theory just with these concepts. Its explanations draw freely upon the full range of physical

Having said this, the fluid approach indicates that the paths employed in the de Broglie-Bohm theory, when shorn of an attached corpuscle, play a key role in quantum mechanics. We can, therefore, invoke the large de Broglie-Bohm-related *oeuvre* as a resource descriptive not of an ensemble of possible tracks of a single particle but of an alternative model of quantum evolution that provides insight not available in other approaches (e.g., the trajectories corresponding to two-slit interference [81, 82]). In particular, the unambiguous solutions given by the de Broglie-Bohm interpretation to problems that have been difficult to solve in other ways can be reinterpreted as answers given by quantum mechanics itself. These tend to be cases requiring definitions of particle-like concepts such as speed and separation that are ambiguous in a pure-wave context. Examples include analyses of tunnelling times (e.g., [83]) and chaos (e.g., [3, 4, 84-94]). In the latter case it is important to note that we can formulate the notion of *Lagrangian* chaos (to be distinguished from the Eulerian variety [95-97]) in which initially adjacent fluid elements diverge exponentially in time. This concept is absent in wave mechanics where, as is well known, the linearity of the wave equation precludes the appearance of (Eulerian) chaos. In contrast, many examples of chaotic behaviour have been found in the (highly nonlinear) Lagrangian description. This may be characterised by positive quantum Lyapounov exponents, the same definition being used for an exponent as for classical dynamical systems [95].

Likewise, what may appear as somewhat arcane unresolved problems in de Broglie-Bohm-like theories – such as whether they admit relativistic formulations – may be recast as open problems within the particle picture of quantum theory.

Finally, we remark on the relation between the Lagrangian picture as presented here and the numerical techniques mentioned in Sec. 1. An important example of the latter is the "quantum trajectory

---

concepts (such as energy and force). This allows a significant improvement in the clarity and consistency with which language is used in quantum theory, where concepts are now no longer just words attached to symbols.

[4] There is debate about where mass is located in the pilot-wave theory [80]. If, as here, the corpuscle is attributed a mass, the fluid model suggests the further question as to what fixes its numerical value.

method" [14]. The idea is to write the Eulerian equations (4.8) and (4.12) in terms of the time derivative along a fluid particle track, $d/dt = \partial/\partial t + v_j\,\partial/\partial x_j$:

$$\frac{d\rho}{dt} = -\rho\frac{\partial v_i}{\partial x_i} \tag{7.1}$$

$$\frac{dS}{dt} = \frac{1}{2m}\left(\frac{\partial S}{\partial x}\right)^2 - V - V_Q. \tag{7.2}$$

The fluid is approximated as a discrete set of particles of mass $m$ and the coupled equations (7.1) and (7.2) are solved together with the particle equation (4.18) to give the variables $x$, $S$ and $\rho$ at a point on a trajectory at time $t$. The solution for the wavefunction evaluated at the location of the $n$th particle is then

$$\psi\big(x_n(t)\big) = \psi_0\big(x_n(0)\big)\exp\left(-\tfrac{1}{2}\int_0^t (\partial v_i/\partial x_i)\big|_{x_n(t)}\,dt\right)\exp\left(\frac{i}{\hbar}\int_0^t L_Q\big(x_n(t)\big)\,dt\right) \tag{7.3}$$

where $L_Q = (1/2m)(\partial S/\partial x)^2 - V - V_Q$ is the Lagrangian of a particle of mass $m$ moving in the potential $V + V_Q$. This and related methods of solution differ from the "full" Lagrangian technique of Secs. 2 and 4 where the Euler-Lagrange equations for the paths are solved independently of the computation of the wavefunction (the latter being deduced from formula (4.15)), and the paths are not populated by particles of mass $m$.

In fact, between the extremes of what we might call the "pure" Lagrangian formalism (in which the wavefunction appears only in the initial conditions) and the "pure" Eulerian formalism (in which the trajectories appear only in extraneous equations), many intermediate pictures, or combinations of equations, are possible. An example similar to that just described was given in Sec. 3 where the Lagrangian equations subject to the initial quasi-potential condition were shown to be equivalent to a set of equations mixing $q$, $S$ and $\rho$. The application of mixtures of equations is well known in numerical hydrodynamics (e.g., [98]).

## 8. Discussion

We have established a formal phase space relation between the wave-mechanical formalism for a single body and a (continuously-)many-particle theory possessing an interaction potential of a certain kind. This is achieved by mapping the particle Hamilton equations together with certain initial conditions into a corresponding Hamiltonian theory that implies the Schrödinger equation together with extraneous field equations. We thereby obtain complementary descriptions of a single quantum process of a quite different kind to the usual notion of complementarity (the application of concepts in mutually incompatible experimental contexts). The two pictures entail different concepts of "quantum state", and consequently tell different stories about the history of a system. Thus, consider a fixed space point $x$ and the passage of time $t:0 \to t$. Then $\psi(x,t)$ describes how the amplitude at the point evolves from $\psi_0(x)$ during the time interval, and similarly for derived quantities such as velocity. On the other hand, for a particle that occupies the point at time $t$, $x = q(a,t)$ describes how the particle travelled there along a unique track through space from an initial point $a$. That is, $\psi$ encodes the temporal history of a space point, while $q$ encodes its "spatial" history. Given the initial wavefunction, the time dependence of either state function can be computed and implies the other. The theory brings the wave equation within the orbit of "rational mechanics" but of course we do not suggest that quantum mechanics can be incorporated in some classical model. Those who would dismiss the trajectory concept when presented in the de Broglie-Bohm context can no longer do so without risk of rejecting quantum mechanics itself.

Whether the exact computational scheme embodied in (2.16) and (4.15) will prove of practical value is uncertain. The little that is known about the solutions to (2.16) comes essentially from studies of the de Broglie-Bohm model. That wave mechanics may be so constructed that it can be mapped into a particle-like theory is not surprising since it is a Hamiltonian system. It is nevertheless notable that it is connected to a deterministic particle system of Newtonian type. The wave-mechanical formulation is undoubtedly simpler but it is arguable that the detailed model provided by the particle approach makes it the more fundamental description [54]. Since the initial particle positions $a$ do not appear in the final computed wavefunction (just as Feynman's paths leave no trace in $\psi$), they invite comparison with, and interpretation as, "hidden variables". Indeed, the step to the de Broglie-Bohm model looks more compelling from this viewpoint but it is necessary to distinguish the de Broglie-Bohm and hydrodynamic interpretations.

We conclude with some further remarks and suggestions for extensions of the theory:

1. Our results constitute an additional reason to formulate hydrodynamical versions of quantum evolution equations. Although hydrodynamics is a natural language in this context, the mathematics is compatible with other theories of continuum mechanics, for example, elasticity. In that case vortex lines may be regarded as (wavefront) dislocations, the circulation (4.14) being a quantized Burgers vector [99, 100].

2. It is straightforward to include an external vector potential in the particle-picture Lagrangian using the standard minimal coupling procedure. The extension to an $N$-body quantum system is also easy in principle – one just extends the range of values taken by the Latin indices. The theory is considerably more complicated, though, as the Jacobian $J$ becomes a sum of products of $3N$ deformation gradients. The fluid analogy also becomes less compelling as a "fluid particle" has $3N$ components, and the Eulerian functions are defined on $3N$-dimensional space.

3. The fluid particle paths may be regarded as analogues in the full wave theory of rays in the geometrical optics limit. The latter (classical) limit is obtained in circumstances where the internal potential and force are negligible in comparison with the external body potential and force, respectively. The result is a fluid-dynamical representation of the classical Hamilton-Jacobi theory (subject to the caveats discussed in [101]). Note that this limit describes a continuous ensemble of non-interacting trajectories moving in the potential $V$, rather than just one.

4. It is anticipated that other field equations of physics will be similarly susceptible to an analysis in terms of fluid paths (e.g., [69]). Indeed, Eulerian pictures for the Pauli, Dirac, Weyl and Maxwell wave equations are well developed [3, 102-104]. However, the equations in the standard Eulerian approach to systems with spin are complicated and difficult to interpret, at least in comparison with the simplicity of the spin 0 Madelung model, and there are ambiguities in identifying suitably defined phases of the relevant wavefunctions (and the associated vortex structures) [105]. It has been suggested that the origin of these problems is that the standard approach works with the "wrong" (angular momentum) representation of the quantum theory, in which the rotational freedoms appear as discrete indices in the wavefunction [3]. The local fluid quantities (density, velocity, spin vector,…) are defined by "averaging" over these indices, and reproducing through them all the information in the wavefunction requires introducing ever more complex quantities and combinations of quantities (especially in the many-particle case; see, e.g., [106]). The alternative procedure advocated in [3] is to start from the angular coordinate representation in which the spin freedoms are represented as continuous parameters $\alpha$ (Euler angles) in the wavefunction, on the same footing as the spatial variables $x$: $\psi(x, \alpha, t)$. This implies a physically clearer and simpler hydrodynamic-like model, in both its Eulerian and Lagrangian guises. The phase $S$ of the wavefunction is immediately identifiable and the equations for $\partial \alpha / \partial t$ and $\partial x / \partial t$ are defined in terms of the gradients of $S$ with respect to the respective coordinates, obvious generalizations of the spin 0 theory. The approach also provides a natural framework to study vortices, which occur in nodal regions in $(x, \alpha)$-space, and the extension to many-body systems is straightforward.

5. The issue of the uniqueness of the Lagrangian model associated with given Eulerian equations has several aspects. For example, without altering the dynamics one can modify the definition of the quantum pressure tensor (2.14) by adding a divergence-free tensor [3, 57]. A further point is whether the Lagrangian (2.4) may be modified so that it is compatible with a law of motion different to that implied by the quasi-potential requirement, in accordance with the alternative guidance equation implied by relativistic constraints [82, 107-109].

6. Although the full translation of the quantum formalism into the Lagrangian language remains to be investigated, some preliminary remarks may be made. For the mean position and momentum in the two pictures we have

$$\langle x_i \rangle = \int x_i \rho(x, t) \, d^3 x = \int q_i(a, t) \rho_0(a) \, d^3 a \tag{8.1}$$

$$\langle p_i \rangle = \int \frac{\partial S(x, t)}{\partial x_i} \rho(x, t) \, d^3 x = \int m \frac{\partial q_i(a, t)}{\partial t} \rho_0(a) \, d^3 a. \tag{8.2}$$

It is clear from these formulas that, as mentioned previously, the fluid coordinates and momenta are playing the role of "Heisenberg $c$-numbers", that is, they carry the time variation while the (Schrödinger) quantum state is static. Although the values of the new variables coincide with the spectra of the position and momentum operators, in themselves they give no clue as to which result will be obtained in a measurement of position or momentum. Their connection with "observables" requires careful consideration, as does the connection between the canonical and unitary mappings. The simultaneous attribution of position and momentum variables to each fluid element does not contradict Heisenberg's relations for reasons already discussed in detail in connection with the de Broglie-Bohm theory [3].

7. The Lagrangian picture exhibits a new quantum symmetry, *viz.* a continuous particle-relabelling transformation with respect to which the Eulerian functions are invariant [68]. This is an analogue of the classical symmetry connected with vorticity conservation [110].

8. We may try to develop the quantum/fluid analogy further and enquire whether the Schrödinger fluid may be modelled in terms of some microscopic structure, much as a classical fluid is ultimately composed of atoms, and if so what this structure must be in order that the bulk or collective behaviour of the "atoms" is governed by Schrödinger's equation. We shall show elsewhere that the Schrödinger fluid may be so modelled if the atoms are governed by a Hamiltonian dynamics involving a particular set of point-like interactions. This throws light on the oft-made claim that the quantum potential acts "nonlocally" in the single-particle case (of course we expect the $N$ "subparticles" that constitute a "fluid particle" in $N$-body quantum theory to interact nonlocally, as is well understood from the de Broglie-Bohm theory [3, 4]).